\def\draftversion{false}

\RequirePackage{ifthen}
\ifthenelse{\equal{\draftversion}{true}}{
  \documentclass[aps,prb,10pt,galley,amsmath,amssymb,showpacs,
                 superscriptaddress]{revtex4}
}{
  \documentclass[aps,prb,10pt,twocolumn,amsmath,amssymb,showpacs,
       longbibliography,superscriptaddress]{revtex4-1}
}

\usepackage{graphicx}
\usepackage[usenames]{color} 
\usepackage{bm} 
\usepackage{dcolumn} 
\usepackage{soul} 
\usepackage{multirow}

\ifthenelse{\equal{\draftversion}{true}}{
  \marginparwidth 2.7in
  \marginparsep 0.5in
  \newcounter{comm} 
  \def\commnext{\stepcounter{comm}}
  \def\commtext{{\bf\color{blue}[\arabic{comm}]}}
  \def\commmar{{\bf\color{blue}[\arabic{comm}]}}
  \def\dvm#1{\commnext\marginpar{\small DV\commmar: #1}\commtext}
  \def\mym#1{\commnext\marginpar{\small MY\commmar: #1}\commtext}
  \def\mlab#1{\marginpar{\small\bf #1}}
  
}{
  \def\dvm#1{}
  \def\mym#1{}
  \def\mlab#1{}
  
}

\def\beq{\begin{equation}}
\def\eeq{\end{equation}}

\def\ABO{$AB$O$_3$}
\def\ABBpO{$A_2BB'$O$_6$}

\def\LNO{LiNbO$_3$}
\def\LOO{LiOsO$_3$}
\def\LTO{LiTaO$_3$}
\def\ZSO{ZnSnO$_3$}
\def\MTO{MnTiO$_3$}
\def\FTO{FeTiO$_3$}
\def\NTO{Ni$_3$TeO$_6$}
\def\LZTO{Li$_2$ZrTeO$_6$}
\def\LHTO{Li$_2$HfTeO$_6$}
\def\MWO{Mn$_3$WO$_6$}

\begin{document}


\title{Domain walls and ferroelectric reversal in corundum derivatives}

\author{Meng Ye}
\email{mengye@physics.rutgers.edu}
\affiliation{
Department of Physics \& Astronomy, Rutgers University,
Piscataway, New Jersey 08854, USA}
\affiliation{
The Institute for Molecular Engineering, The University of Chicago,
Chicago, Illinois 60637, USA}

\author{David Vanderbilt}
\affiliation{
Department of Physics \& Astronomy, Rutgers University,
Piscataway, New Jersey 08854, USA}

\date{\today}

\begin{abstract}

Domain walls are the topological defects that mediate polarization
reversal in ferroelectrics, and they may exhibit quite different
geometric and electronic structures compared to the bulk. Therefore,
a detailed atomic-scale understanding of the static and dynamic
properties of domain walls is of pressing interest. In this work, we
use first-principles methods to study the structures of $180^{\circ}$
domain walls, both in their relaxed state and along the ferroelectric
reversal pathway, in ferroelectrics belonging to the family of corundum
derivatives. Our calculations predict their orientation, formation
energy, and migration energy, and also identify important couplings
between polarization, magnetization, and chirality at the domain walls.
Finally, we point out a strong empirical correlation between the height
of the domain-wall mediated polarization reversal barrier and the local
bonding environment of the mobile $A$ cations as measured by bond
valence sums. Our results thus provide both theoretical and empirical
guidance to future searches for ferroelectric candidates in materials of
the corundum derivative family.

\end{abstract}

\maketitle


\section{Introduction}
Ferroelectrics (FEs) are materials with a spontaneous electric
polarization that can be reversed by an external electric field.
\cite{07Book-Rabe} Because the polarization in FEs couples to strains
and temperature gradients as well as electric fields, FEs have broad
potential applications and are already commonly used in sensor
\cite{00FEsensor} and memory \cite{00Book-Scott} applications.
In these materials, regions of different polarization orientation,
known as FE domains, often coexist, and the interface between two
FE domains is referred to as a domain wall (DW). The hysteresis
observed during the switching of FE materials is caused by the
nucleation and growth or shrinkage of domains through the motion
of DWs in response to an applied electric field favoring one type
of domain over the other.  To complicate matters, defects can be
attracted to the DW and contribute to pinning of the DW motion.
\cite{03He-defectDW}

DWs can be seen as topological defects which have different geometric
and electronic structures compared to the bulk, so DWs may exhibit rich
physics that is not present in the bulk. For instance, in BiFeO$_3$,
experiments have shown that the DWs behave as conductive channels
in the otherwise insulating background. \cite{09BFODWconduct}
The same DWs are suggested to exhibit photovoltaic properties as well.
\cite{10BFODWPV} Furthermore, it has recently been observed that
charged FE DWs, which are energetically unfavorable in general, are
abundant in hybrid improper FEs $\rm{(Ca,Sr)_3Ti_2O_7}$. \cite{15chargedDW}
Moreover, in hexagonal manganite YMnO$_3$, the FE DWs form
topologically protected vortices, \cite{10YMnO3FEvortex} and
alternating magnetic moments are found at the FE DW around the
vortex core. \cite{12RMnOMFvortex} In addition, the FE DWs are also
observed interlocking with chiral DWs in \NTO. \cite{15NTOchiralFEDW}

Recently, attention has been drawn to a family of polar materials that
can be regarded as derivatives of the corundum $A_2$O$_3$ structure
but with cation ordering on the $A$ site. The best-known examples are
the binary-cation materials LiNbO$_3$ and LiTaO$_3$, belonging to
the general formula \ABO, but ternary-cation \ABBpO\ are also under
investigation.\cite{MY16-FEcd} However, polar insulators are not
necessarily FE; they can only be considered as such if their polarization
can be reversed by an electric field.  For example, wurtzite oxides
such as ZnO are polar, but undergo dielectric breakdown well before
polarization reversal can occur.  Unfortunately, many of the newly
proposed \ABO\ and \ABBpO\ materials are the result of high-pressure
syntheses and are only available as powder samples, so that it remains
unclear whether or not they would be FE in single-crystal or thin-film form.

Thus, until the experimental growth of single crystals and subsequent
observation of domain wall motion can be achieved, theory can play an
important role in identifying which materials are most likely to be
switchable. We previously carried out first-principles calculations of
coherent bulk switching in this class of materials, \cite{MY16-FEcd}
but did not consider DW-mediated mechanisms. As the DW structure
is very different from the bulk, it is interesting to investigate
how the DW structure would influence the reversal barrier. The physics
of DWs in the corundum derivatives is especially rich because some of
these materials are chiral, some are magnetically ordered, and some are
both.  As a result, the distinct structure at the DW can have additional
degrees of freedom that may lead to unique interactions between chiral
and FE domains and magnetoelectric effects.

In this work, we use first-principles methods to study the formation and
motion of FE DWs at the atomic scale in order to characterize their
intrinsic properties and their role in the FE reversal process. The FE
candidates that we consider are \LNO (LNO), \LTO, \ZSO, \FTO, \MTO,
\LZTO, \LHTO, and \MWO. Our study of the $180^{\circ}$ charge-neutral
DWs predicts not only the energy and preferred orientation of static DWs,
but also the migration path and energy barrier for displacement of a
domain wall by a lattice vector during domain reversal. In the case of
ferrimagnetic \MWO, our DW formation energy results also suggest
that FE DWs are simultaneously magnetic DWs, potentially giving rise
to a strong extrinsic magnetoelectric coupling through DW motion.
Finally, our results allow us to clarify how the DW-mediated reversal
barrier is strongly correlated with the local bonding environment of
the $A$ cations as measured by bond valence sums.

The paper is organized as follows.
Sec.~\ref{sec:methods} presents a brief summary of our first-principles
methods. In Sec.~\ref{sec:results}A we explain the principles and
assumptions behind our construction of DWs, while the structures
and energies of DWs with different orientations are discussed in
Sec.~\ref{sec:results}B. In Sec.~\ref{sec:results}C we consider
different magnetic structures at the DWs, leading to the discovery of
a potential DW-mediated magnetoelectric coupling in ferrimagnetic \MWO.
Finally, in Sec.~\ref{sec:results}D we use a structural constraint method
to study the migration path and barrier energy in the DW-mediated
polarization reversal process, and find a strong correlation between the
barrier energy and the local bonding environment of the $A$ cations.
A brief summary is given in Sec.~\ref{sec:summary}.

\section{First-principles methods}\label{sec:methods}
The calculations are performed with plane-wave density functional theory
(DFT) implemented in VASP \cite{VASP} with PBEsol \cite{PBEsol} as
the exchange-correlation functional. The ionic core environment is
simulated by projector augmented-wave (PAW) pseudopotentials. \cite{PAW}
We use a Hubbard $U=4.2$\,eV on the 3$d$ orbitals of Mn and Fe. \cite{LDAU}
The magnetic moments are collinear and spin-orbit coupling is neglected.
The cutoff energy for all calculations is 550\,eV. The energy error
threshold varies slightly in different calculations, but an accuracy
between $1.0\times10^{-5}$ and $1.0\times10^{-7}$\,eV is achieved.
The forces are reduced below 0.01 eV/\AA\ in the DW structural
relaxations.
2$\times$6$\times2$ and 6$\times$6$\times$1 $k$ meshes are
used in the calculations of X-walls and Y-walls respectively
as defined below.

\section{Results and discussion}\label{sec:results}

\subsection{Construction of domain walls}
\begin{figure}
\begin{center}
  \includegraphics[width=8cm]{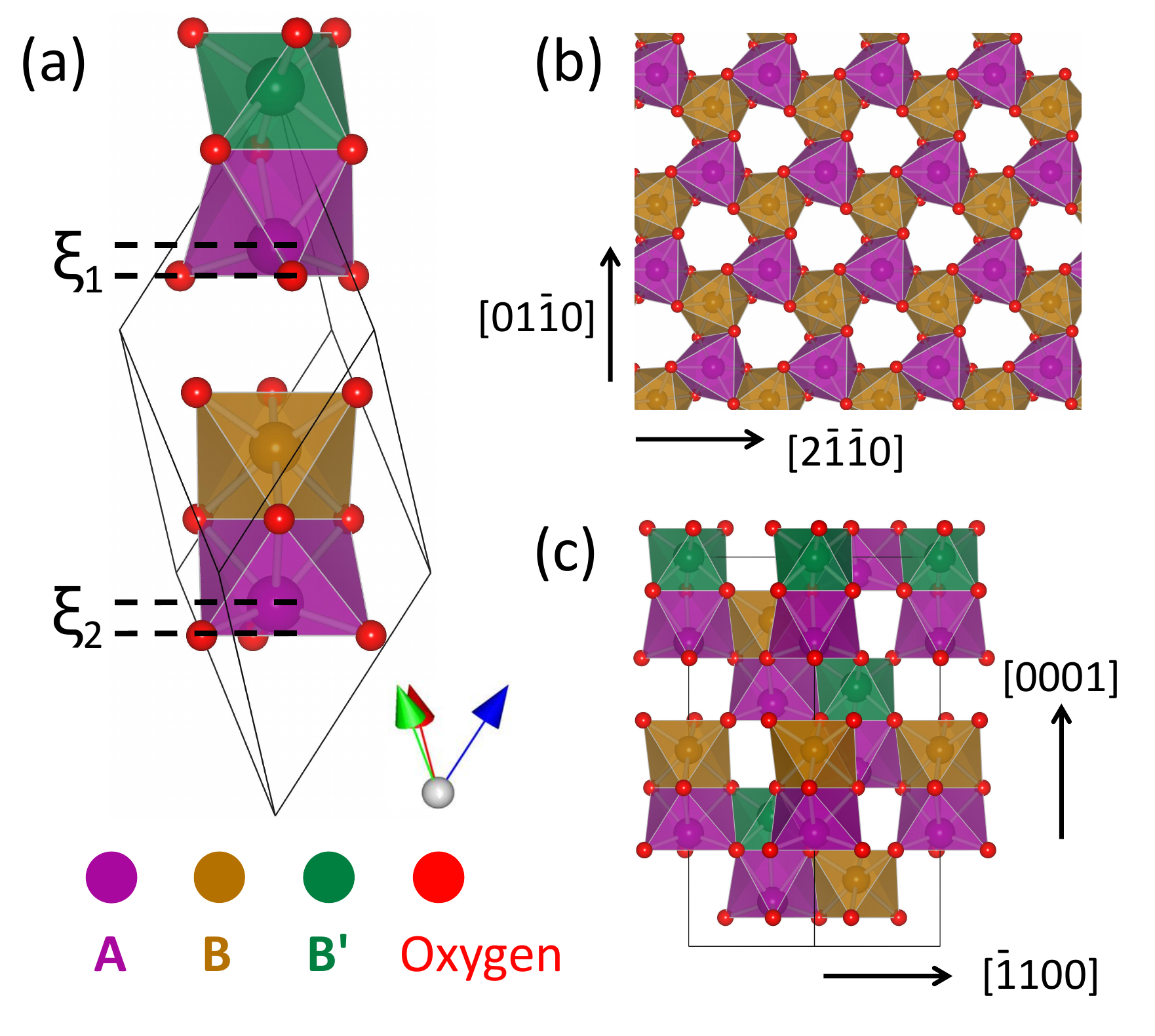}
    \caption{Structure of ordered-LNO corundum derivatives \ABBpO,
    and of LNO-type \ABO\  corundum derivatives if $B'\!=\!B$.
    (a) Side view
    of the rhombohedral unit cell. $\xi_1$ (or $\xi_2$) is the vertical
    distance between an $A$ cation and the oxygen plane that it
    penetrates during the polarization reversal. (b) Top view of the
    $AB$ layer and (c) side view in the enlarged hexagonal-setting cell.}
    \label{fig:GStopside}
\end{center}
\end{figure}

The general structure of corundum derivatives \ABO\ and \ABBpO\ were
introduced previously.\cite{MY16-FEcd} Here we only focus on the
\LNO\ (LNO)-type and ordered-LNO structures, which are compatible with
ferroelectricity along the rhombohedral axis. In Fig.~\ref{fig:GStopside},
both the rhombohedral unit cell and views from different hexagonal
directions are illustrated. Each cation is in a distorted oxygen
octahedron; these only fill two thirds of the octahedral sites, leaving
cation-vacant positions that we denote by ``$-$". The FE reversal is
driven by migration of each $A$ cation from its own oxygen octahedron
to the cation-vacant octahedron above or beneath it. \cite{LiNb/TaO-cal,LiNb/TaO-cal2}
The reversal path can be qualitatively described by two variables
$\xi_1$ and $\xi_2$ defined as the vertical distances between each $A$
cation and the oxygen plane that it penetrates during the polarization
reversal. \cite{MY16-FEcd} The definitions of $\xi_1$ and $\xi_2$ are
also shown in Fig.~\ref{fig:GStopside}(a).

In order to study the properties of DWs, we construct a supercell with
a polarization-up domain and a polarization-down domain that are
related by inversion through a center located in the FE DW separating
them. Because of periodic boundary conditions, there are always two
DWs in the supercell, one denoted as DW$_{\Uparrow\Downarrow}$
(polarization up on the left and down on the right) and the other as
DW$_{\Downarrow\Uparrow}$ (the opposite case).  In corundum
derivatives the R3c symmetry of the LNO-type structure ensures that
the DW$_{\Uparrow\Downarrow}$ and DW$_{\Downarrow\Uparrow}$
walls are equivalent, but they become inequivalent in the ordered-LNO
structure with R3 symmetry. Because DW$_{\Uparrow\Downarrow}$
and DW$_{\Downarrow\Uparrow}$ always come in pairs in our
supercell calculations, we report only the average formation energy of
DW$_{\Uparrow\Downarrow}$ and DW$_{\Downarrow\Uparrow}$ for
the ordered-LNO materials.

\begin{figure}
\begin{center}
  \includegraphics[width=5cm]{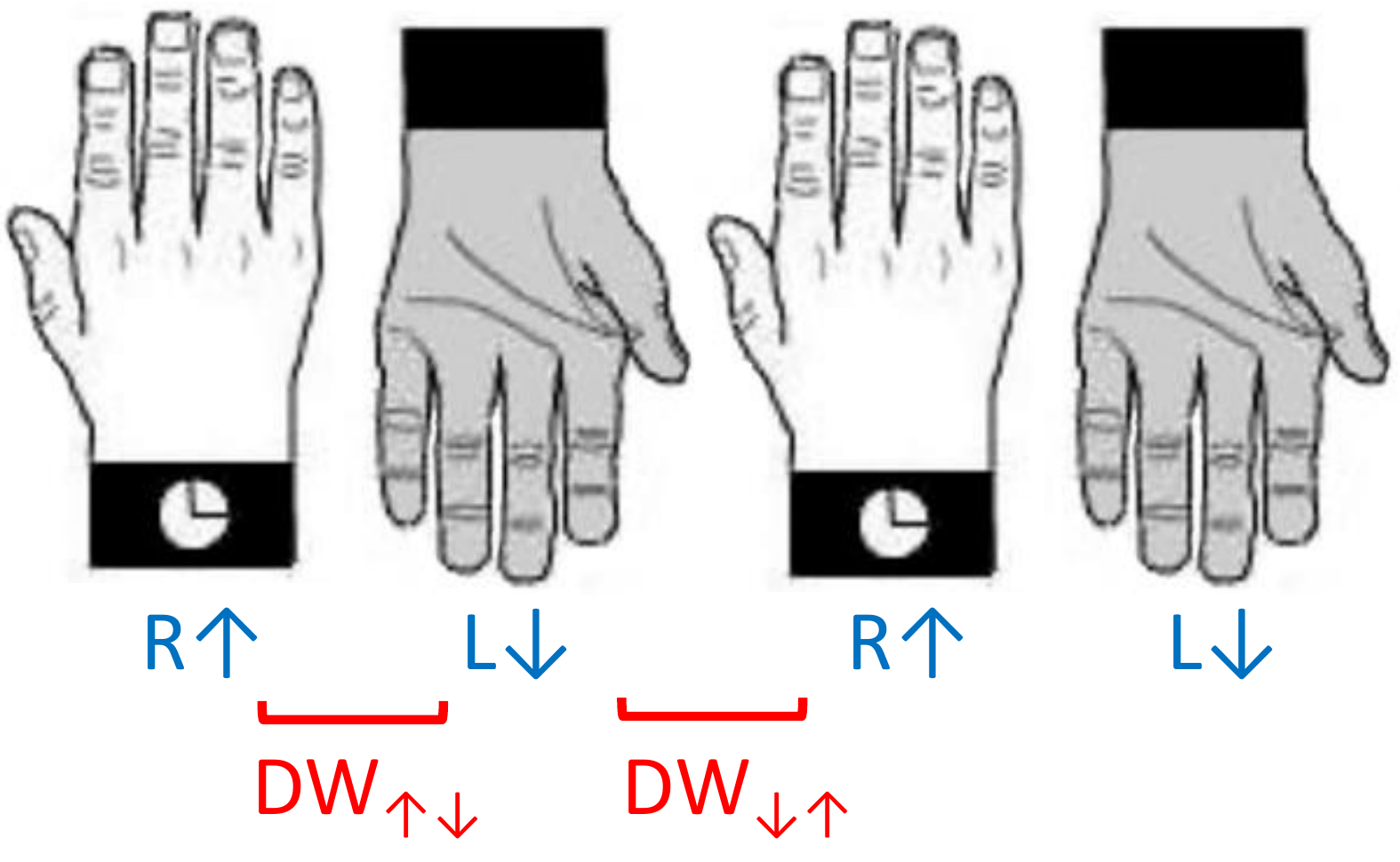}
    \caption{Illustration of domains and DWs in ordered-LNO materials.
    FE DWs are formed between (R$\Uparrow$) domains and (L$\Downarrow$)
    domains. The DW between adjacent thumbs represents DW$_{\Downarrow\Uparrow}$
    and the DW between adjacent little fingers represents  DW$_{\Uparrow\Downarrow}$.
    Left and right hands represent left (L) and right (R) chirality, and
    the direction in which the fingers point ($\Uparrow$ or $\Downarrow$)
    represents the polarization direction.}
    \label{chiral}
\end{center}
\end{figure}

The intrinsic difference between DW$_{\Uparrow\Downarrow}$ and
DW$_{\Downarrow\Uparrow}$ in the ordered-LNO structure arised
from the chiral nature of the structure. The ordered-LNO space group
is R3, which does not contain any mirror symmetries and is therefore
chiral. In contrast, the LNO-type structure is not chiral because of its
R3c symmetry. The hands shown in Fig.~\ref{chiral} illustrate how the
chirality of the FE domains implies that DW$_{\Uparrow\Downarrow}$
and DW$_{\Downarrow\Uparrow}$ are inequivalent. The
fingers point in the polarization direction, and the left (L) or right (R)
hand represents the L or R chirality. The ferroelectricity in corundum
derivatives is generated by breaking the inversion symmetry from the
reference structure with symmetry R$\bar{3}$, \cite{MY16-FEcd} so
the two FE domains with opposite polarization are related by an
inversion operation. Since the inversion operation flips both polarization
and chirality, the FE DW is also a chiral DW. Such an interlocking effect
between polarization and chirality at the DW was observed in chiral
pyroelectric \NTO. \cite{15NTOchiralFEDW} It is also clear that there
are two kinds of interfaces between hands as shown in Fig.~\ref{chiral},
one between thumbs and the other one between little fingers, and these
correspond to the DW$_{\Downarrow\Uparrow}$ and DW$_{\Uparrow\Downarrow}$.

The presence of this chirality can also help explain the observed shape
of domains in corundum-derivative materials. In \LNO\ and other
LNO-type materials with R3c symmetry, the Wulff construction \cite{Wulff}
implies that the ideal domain shape is a regular hexagon, because
DW$_{\Uparrow\Downarrow}$ and DW$_{\Downarrow\Uparrow}$
have identical formation energies. As the asymmetry between the
energies of  DW$_{\Uparrow\Downarrow}$ and DW$_{\Downarrow\Uparrow}$
increases, the Wulff shape first distorts into an equiangular hexagon
having alternating long and short sides, and eventually into a triangle
in which only the favored DW is exposed. In fact, regular hexagonal
domains are observed in \LNO \cite{LiNbO-DWexpt} while triangular
domains are found in \NTO\ with the ordered-LNO structure. \cite{15NTOchiralFEDW}

To arrive at our DW configurations, we assume that the $B$/$B'$ and O
sublattices are preserved throughout the supercell, so that the DW
only results from the interchange of $A$ and $-$ sublattices (that is,
migration of $A$ cations into vacancies) on one side of the DW. This is
motivated by the greater mobility of the $A$ cation species. In addition,
only the $180^{\circ}$ charge-neutral DW is considered, in which the
polarization direction is parallel to the DW plane but antiparallel
between domains.

\subsection{Orientation of domain walls}
\begin{figure}
\begin{center}
  \includegraphics[width=8.5cm]{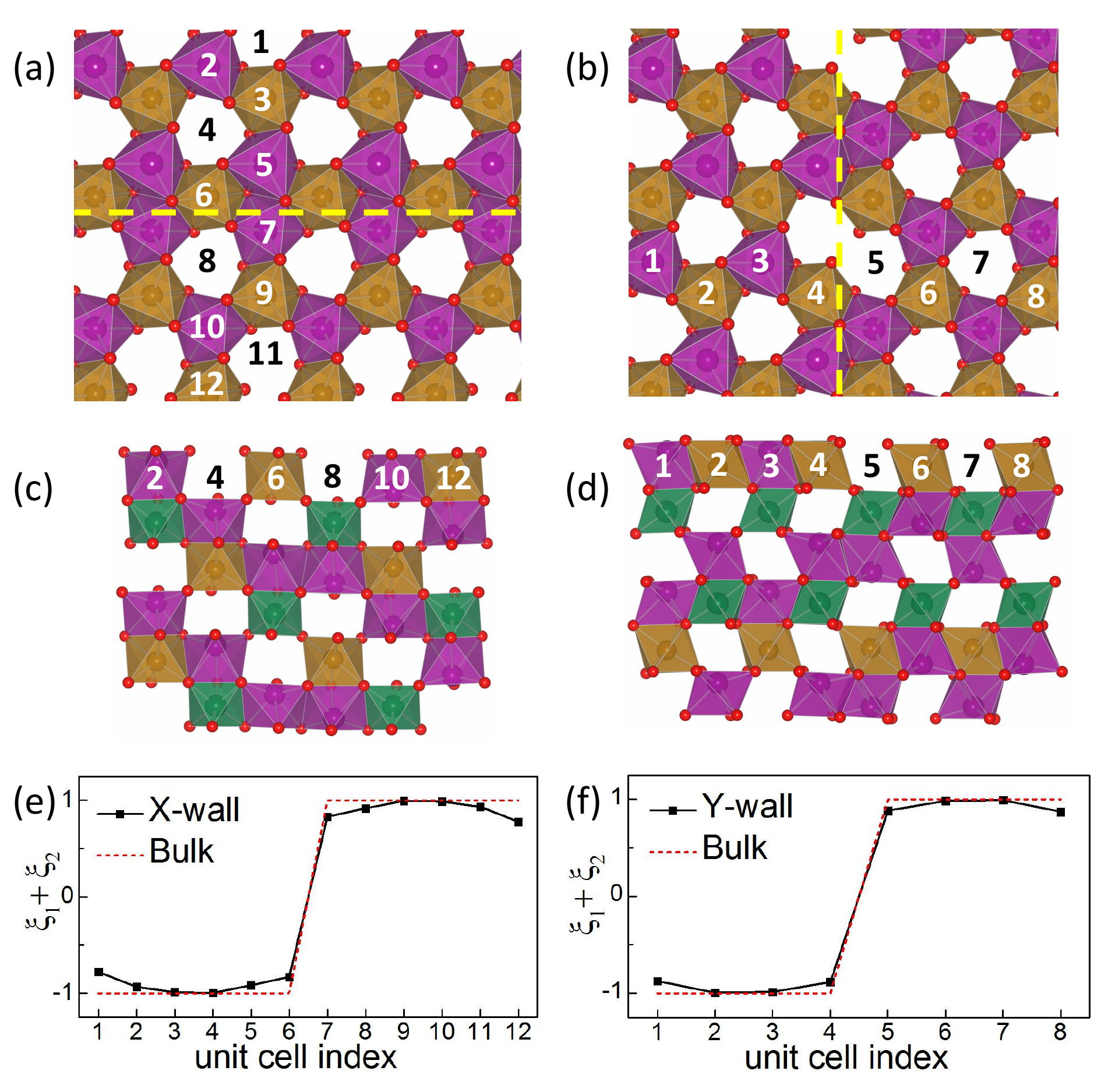}
    \caption{Structures of the X-wall in the 6+6 supercell (left)
    and the Y-wall in the 4+4 supercell (right).
    (a,b) Top views of the X-wall and Y-wall. The number in each
    octahedron is the unit cell label. The X-wall lies in the $x$-$z$ or
    (01$\bar{1}$0) plane and is located between the 6th and the 7th
    unit cell as shown by the dashed yellow line. The Y-wall lies in the $y$-$z$
    or (2$\bar{1}\bar{1}$0) plane and is located between the 4th and
    the 5th unit cell.
    (c,d) Side views of the X-wall and Y-wall. Odd-number cells are
    behind even-number cells in the X-wall.
    (e,f) The $\xi_1+\xi_2$ displacement profile of an X-wall and a Y-wall.}
    \label{fig:XYDW}
\end{center}
\end{figure}

Since corundum derivatives have three-fold symmetry, there are two
types of $180^{\circ}$ DWs depending on the orientation of the DW
plane. We refer to a DW in the $x$-$z$ plane as an X-wall and one in
the $y$-$z$ plane as a Y-wall. Top and side views
of the X-wall and Y-wall are shown in Fig.~\ref{fig:XYDW} in comparison
with the bulk structure in Fig.~\ref{fig:GStopside}(b-c). In the layer
shown in Fig.~\ref{fig:XYDW}(a), octahedra containing $A$ cations are
densely packed at the X-wall. However, there are also layers where the
octahedra at the X-wall are all cation-vacant. In short, the X-wall
consists of alternating dense and sparse octahedral layers. In comparison,
the $A$ and $-$ sublattices are more evenly spaced at the Y-wall.

To calculate the formation energies, we construct 6+6 and 6+7 supercells
for the X-wall, and 3+4, 4+4, and 4+5 supercells for the Y-wall. Here the
supercell notation $m+n$ means that $m$ unit cells are of polarization
down and $n$ unit cells are of polarization up. The $m+m$ supercell
preserves some symmetry, while the $m+(m+1)$ supercell has none
because of the asymmetry of the size of up and down domains. In
the $m+m$ supercells, the up and down domains are related by an
inversion symmetry through a center lying in the DW. Furthermore, for
the LNO-type structure, the identical DW$_{\Uparrow\Downarrow}$ and
DW$_{\Downarrow\Uparrow}$ are also related by two-fold rotation.
The computed relaxed displacements $\xi_1 + \xi_2$ in each cell of
the 6+6 X-wall and the 4+4 Y-wall of \LZTO\ are shown
in Fig.~\ref{fig:XYDW}(e) and (f).
The displacement profiles suggest that the DWs
in corundum derivatives are atomically sharp, similar to the
DWs of perovskites. \cite{96BTODW,02PTODW} Meanwhile, our calculations
also predict that the Y-wall is energetically favored in all the cases
that we have studied, as shown by the converged DW formation energies
in Table~\ref{tab:XYDW}. Our results for \LNO\ and \LTO\ are consistent
with the earlier DW simulations. \cite{10LNODWcal} Experimental
observations on the domains of \LNO\ also confirm that the Y-wall is
more favorable. \cite{LiNbO-DWexpt}

\begin{table}
\begin{center}
\begin{ruledtabular}
    \caption{Formation energies of X-walls and Y-walls, in units of mJ/m$^2$.
    For the ordered-LNO structure, the formation energy is averaged
    between the DW$_{\Uparrow\Downarrow}$ and DW$_{\Downarrow\Uparrow}$.
    The magnetic orderings are shown in the column labeled ``Mag.". }
    \begin{tabular}{ccccccccc}
    \multicolumn{1}{c}{LNO-type}&\multicolumn{1}{c}{Mag.}&\multicolumn{1}{c}{X} &\multicolumn{1}{c}{Y} &&
    \multicolumn{1}{c}{Ordered-LNO}&\multicolumn{1}{c}{Mag.}&\multicolumn{1}{c}{X} &\multicolumn{1}{c}{Y}\\ \hline
    \LTO &        &71  &63  &&\LZTO&          &29 &20 \\
    \LNO &        &160 &138 &&\LHTO&          &30 &21 \\
    \ZSO &        &106 &81  &&\MWO &$uud$-$uud$ &68 &42 \\
    \MTO &$ud$-$ud$ &171 &153 &&\MWO &$udu$-$dud$ &67 &41 \\
    \FTO &$ud$-$ud$ &183 &108 &&\MWO &$udu$-$udu$ &75 &45  \\
    \end{tabular}\label{tab:XYDW}
\end{ruledtabular}
\end{center}
\end{table}

\subsection{Magnetic and magnetoelectric domain walls}
Some corundum derivatives are magnetic compounds and exhibit magnetic
order. Here we use notations like ``$udu$'' to represent spin-up ($u$) and
spin-down ($d$) states on magnetic cations $A_1$, $A_2$ and $B$ in that
order, where $A_1$ and $A_2$ are face-sharing with $B'$ and $B$ cations
respectively. Since spin-orbit coupling is neglected in our calculations,
``up" and ``down" are not necessarily $\pm\hat z$. In our previous work
\cite{MY16-FEcd} it was shown if the magnetic order is constrained
to preserve the rhombohedral unit cell, the ground state is antiferromagnetic
(or $ud$) for \MTO\ and \FTO, while in \MWO\ the lowest-energy state
is ferrimagnetic $udu$ but with $uud$ close in energy.

\begin{figure}
\begin{center}
  \includegraphics[width=8.5cm]{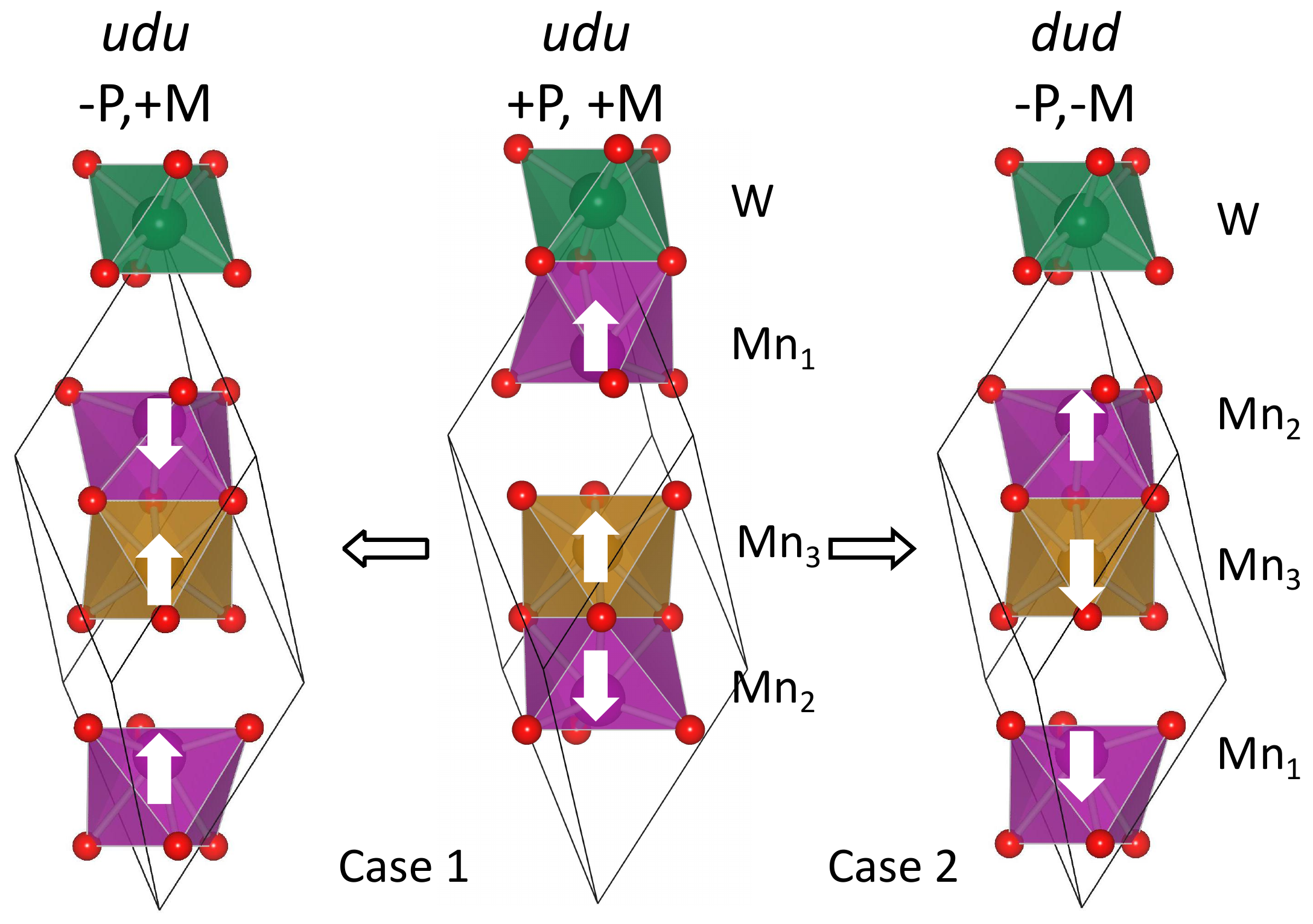}
    \caption{Two possible magnetic orders at FE DWs in \MWO. The
    structure in the center has polarization and magnetization
    ($+$P,$+$M) with the magnetic order $udu$. The structures on the
    left and right both have polarization $-$P but the left one has
    the magnetic order $udu$ while the right one is $dud$. In Case 1,
    the FE DW is formed between structures in the center and on the
    left. In Case 2, the FE DW is formed by the central and rightward
    structures.}
    \label{fig:MEDW}
\end{center}
\end{figure}

Because of the time-reversal symmetry, a global reversal of all the
spins does not affect the total energy, e.g., $udu$ and $dud$ have
exactly the same energy in the bulk of ferrimagnetic compounds such
as \MWO. This means that there are two possibilities for the magnetic
order across a FE DW: the magnetization may be the same on both
sides of the DW, or it may reverse.  The former case is illustrated by
Case 1 in Fig.~\ref{fig:MEDW}, and the corresponding DW is denoted
by $udu$-$udu$. Here the letters before and after ``-" represent
magnetic orders in two neighboring domains. If the magnetic order
reverse across the DW, as shown in Case 2 of the figure, the DW
is denoted by $udu$-$dud$. Therefore, the $udu$-$dud$ FE DW is
also a magnetic DW. Similarly, there are also $ud$-$ud$ DW and
$ud$-$du$ DWs for \MTO\ and \FTO, but neither of them has a net
magnetization.

 \begin{figure}[b]
\begin{center}
  \includegraphics[width=8.5cm]{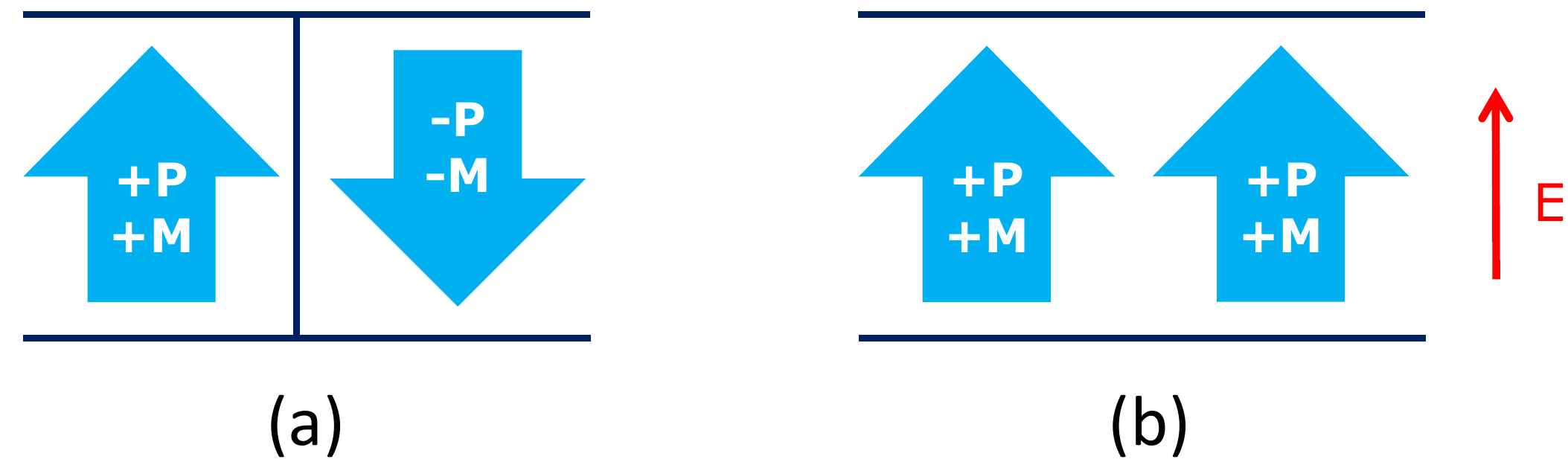}
    \caption{Illustration of magnetoelectric coupling at FE DWs
    when $udu$-$dud$ is favored.
    (a) Magnetization direction is coupled to the polarization direction
    in different FE domains. (b) When the polarization is reversed by
    an electric field, the magnetization reverses as well.}
    \label{fig:MEDWreversal}
\end{center}
\end{figure}

Our computed formation energies for both the $udu$-$udu$ and
$udu$-$dud$ DWs are summarized in Table.~\ref{tab:XYDW}.
Interestingly, the results suggest that the magnetization-reversing
$udu$-$dud$ DW is energetically favored, implying that the FE and
magnetic domains are interlocked. This has interesting consequences.
When a switching electric field is applied to a polydomain FE sample,
the FE DWs move so that the favored domains grow at the expense
of the unfavored ones, as illustrated in Fig.~\ref{fig:MEDWreversal}.
Because the magnetic order reverses across the DW, the magnetization
is simultaneously switched in the same process.  This provides an
extrinsic mechanism for magnetoelectric coupling that could have
useful device implications. Other similar cases of DW-mediated
interlocking of two order parameters have been discussed previously.
\cite{11Benedek-HybridImproper}

As the spin-orbit coupling is not included in our calculations, the
origin of the coupling between magnetization and polarization at
the DW should be categorized as an exchange-striction effect.
Comparing the bulk and DW structures, it is found that, for example,
the $A$ octahedron (purple color) is only edge sharing with three $B$
octahedra in Fig.~\ref{fig:GStopside}(b), but at the Y-wall shown
in Fig.~\ref{fig:XYDW}(d), the $A$ octahedron has an additional
edge-sharing $A$ octahedron at the DW. Therefore, the different
neighbour environments lead to different exchange energies at the
two kinds of DW, explaining the interlocking of magnetism and
 polarization.

In the above calculations, the magnetic DW is assumed to be as sharp
as the FE DW.  One may recall that in most magnetic materials the
magnetic DWs are much broader, and wonder whether we should consider
a broad magnetic DW instead.  We don't believe so.
In most ferromagnetic DWs, the exchange energy prefers a gradual change
of spin directions at the DW, but the magnetic anisotropy favors an
abrupt change of spin directions at the DW. Thus, in the case of strong
exchange and weak spin anisotropy, magnetic DWs are much broader than FE
DWs. However, in $udu$ \MWO, because of the distinct structure at the
FE DW, the exchange energy also prefers the spins to align oppositely
($udu$-$dud$) across the DW. Therefore, both exchange energy and
anisotropy support the sharp magnetization change at the FE DW.

\subsection{Domain wall reversal}
\begin{figure}
\begin{center}
  \includegraphics[width=8.5cm]{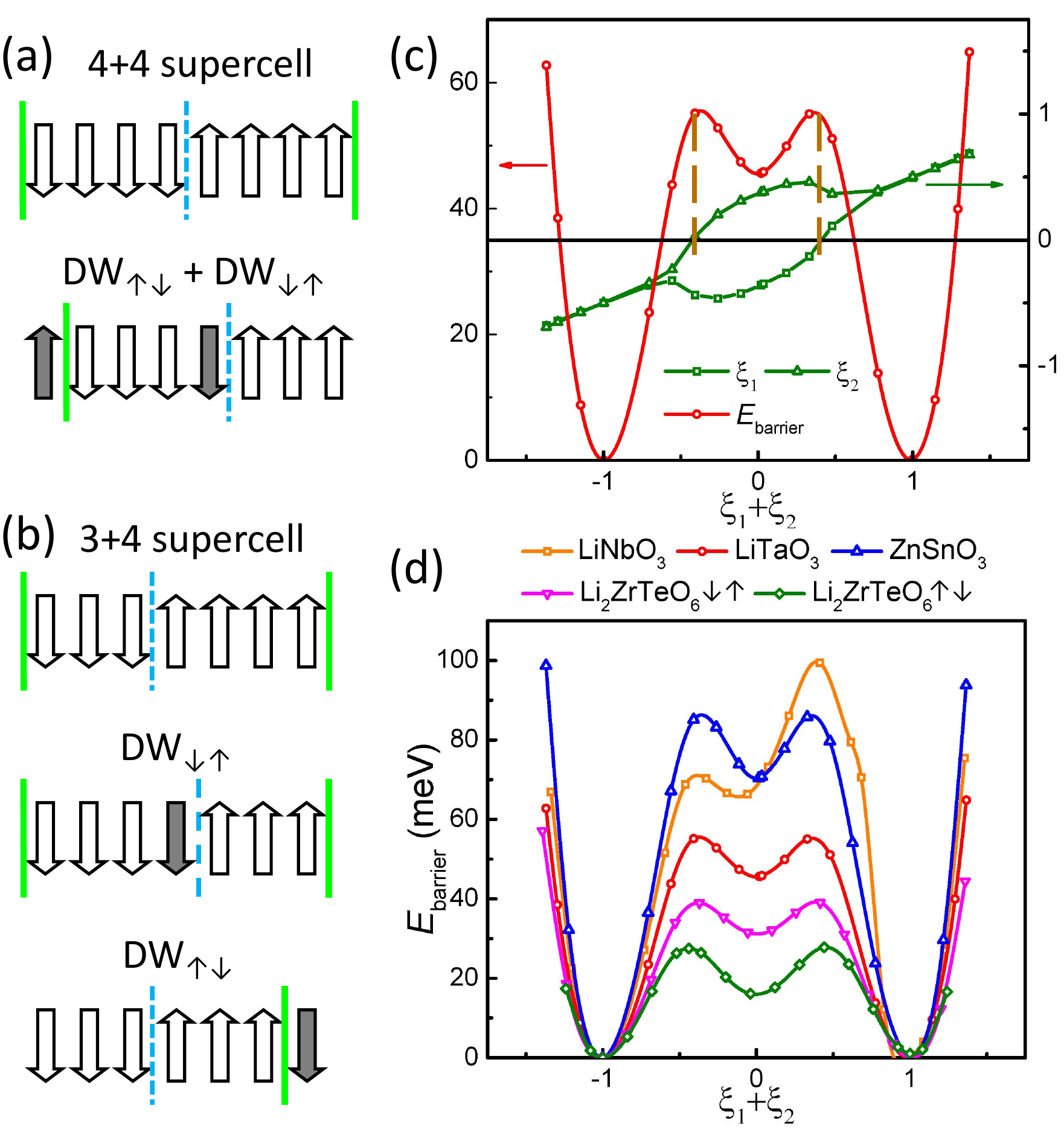}
    \caption{DW-mediated FE reversal in corundum derivatives.
    (a-b) Illustrations of DW motions in 4+4 and 3+4 supercells.
    The upward and downward arrows represent the polarization in each
    unit cell. The dashed blue line represents the DW$_{\Downarrow\Uparrow}$
    and the solid green line is the DW$_{\Uparrow\Downarrow}$. The filled
    black arrows represent the polarization that is reversed during the
    DW motion.
    (c) Energy profile of the DW reversal in \LTO\ and the evolution of
    $\xi_1$ and $\xi_2$. The dashed brown lines highlight the positions
    where $\xi_1=0$ and $\xi_2=0$.
    (d) Energy profiles of the DW reversal for selected corundum
    derivatives. The results of both the DW$_{\Downarrow\Uparrow}$ and
    DW$_{\Uparrow\Downarrow}$ are included for \LZTO. Energy units are
    meV per unit cell.
    }\label{fig:DWreversal}
\end{center}
\end{figure}

The polarization reversal at the DW is associated with DW motion. For
instance, in the 4+4 supercell illustrated in Fig.~\ref{fig:DWreversal}(a),
the simultaneous polarization reversal at the 1st and the 5th cells
(black arrows) causes the DW$_{\Uparrow\Downarrow}$ and
DW$_{\Downarrow\Uparrow}$ to move to the right by one unit cell.
Similarly, in the 3+4 supercell shown in Fig.~\ref{fig:DWreversal}(b),
the polarization reversal at the 4th cell is accompanied by the rightward
motion of the DW$_{\Downarrow\Uparrow}$, and the polarization
reversal at the 7th cell leads to the leftward motion of the
DW$_{\Uparrow\Downarrow}$. In order to insure that the supercells
before and after the DW displacement are equivalent, the $m+m$
supercell always involves the motion of both DW$_{\Uparrow\Downarrow}$
and DW$_{\Downarrow\Uparrow}$. By contrast, the $m+(m+1)$
supercell allows just one selected DW to move. Therefore, while either
type of supercell can be used to study the barriers for DW motion in
LNO-type materials, only the $m+(m+1)$ supercells are suitable for
obtaining the corresponding information about DW$_{\Uparrow\Downarrow}$
and DW$_{\Downarrow\Uparrow}$ walls separately in the ordered-LNO
case.

The adiabatic polarization reversal at the DW is achieved by using the
reaction coordinate $\xi_1+\xi_2$ as a structural constraint \cite{MY16-FEcd}
and applying it only to the unit cell at the DW. The energy profiles
of the DW-mediated reversal of selected materials are illustrated in
Fig.~\ref{fig:DWreversal}(d), and the reversal barriers are listed in
Table.~\ref{tab:DWbarrier}. This DW-mediated barrier is much
lower in energy than the coherent reversal barrier reported in our
previous work.\cite{MY16-FEcd} For instance, the coherent
barrier of \LTO\ is 129\,meV while the DW-mediated barrier is
only 55\,meV. This huge energy reduction is caused by the distinct
structure at the DW. Since the symmetry at the DW is much lower
than in the bulk, there are more phonon modes, e.g., the breathing
modes of the oxygen triangle, that can undergo displacement to
lower the energy barrier. In addition, the DW-mediated reversal
paths are insulating in all the cases that we have calculated including
that of \FTO, which we reported to have a conducting coherent
reversal path in our previous calculations. \cite{MY16-FEcd}

The energy profiles shown in Fig.~\ref{fig:DWreversal}(d) are
symmetric with respect to $\xi_1+\xi_2=0$ for most candidates,
as their DW structures have inversion symmetry at $\xi_1+\xi_2=0$.
The only asymmetric profile in Fig.~\ref{fig:DWreversal}(d) is that
of \LNO. This is cause by an in-plane unstable polar mode at the
midpoint structure, which breaks the local inversion symmetry at
$\xi_1+\xi_2=0$. This unstable E$_u$ mode in \LNO\ has been
previously reported in the literature. \cite{MY16-FEcd,LiNb/TaO-cal2}

\begin{table}
\begin{center}
\begin{ruledtabular}
    \caption{DW-mediated polarization reversal barriers $E_{\rm{barrier}}$ for
    corundum derivatives. The energy barriers of DW$_{\Uparrow\Downarrow}$ and
    DW$_{\Downarrow\Uparrow}$ are the same in the LNO-type structure,
    but different in the ordered-LNO structure. Energy units are meV per
    unit cell.}
    \begin{tabular}{lccclccc}
    \multicolumn{1}{c}{LNO-type}   &\multicolumn{1}{c}{Mag.}&\multicolumn{1}{c}{$\Uparrow\Downarrow=\Downarrow\Uparrow$}  &&
    \multicolumn{1}{c}{Ordered LNO}&\multicolumn{1}{c}{Mag.}&\multicolumn{1}{c}{$\Uparrow\Downarrow$} &\multicolumn{1}{c}{$\Downarrow\Uparrow$}\\ \hline
    \LTO  &          &55  &&\LZTO &            &28 &39   \\
    \LNO  &          &98  &&\LHTO &            &32 &42   \\
    \ZSO  &          &86  &&\MWO  &$uud$-$uud$ &210 &161 \\
    \MTO  &$ud$-$ud$ &229 &&\MWO  &$udu$-$dud$ &212 &181 \\
    \FTO  &$ud$-$ud$ &394 &&\MWO  &$udu$-$udu$ &207 &175 \\
    \end{tabular}\label{tab:DWbarrier}
\end{ruledtabular}
\end{center}
\end{table}

In Fig.~\ref{fig:DWreversal}(c), we use the results for \LTO\ as an example
to clarify the relationship between the energy profile and the evolution
of $\xi_1$ (or $\xi_2$) at the DW. Similar to the results in our previous
work,\cite{MY16-FEcd} $\xi_1\neq\xi_2$ when the reaction coordinate
$\xi_1+\xi_2$ approaches zero, which means that the two $A$ cations do
not migrate simultaneously. However, the barrier structures for the DW-mediated
reversal are qualitatively different from those for the coherent
reversal, where the energy profile displayed a single maximum at $\xi_1+\xi_2=0$
in most cases. \cite{MY16-FEcd} In contrast, the DW-mediated energy profile
has two energy maxima located at approximately $\xi_1=0$ and $\xi_2=0$,
as highlighted by the dashed vertical lines in Fig.~\ref{fig:DWreversal}(c).
Each of these configurations corresponds to the moment when one of the
$A$ cations passes through an oxygen plane.

For magnetic compounds, their DW motions may be accompanied by
spin flips at the DWs. In the reversal process, the $A_1$ cation
migrates away from the $B'$ cation and becomes face sharing with
the $B$ cation. Similarly, the $A_2$ cation moves away from
the $B$ cation and forms face-sharing octahedra with the $B'$
cation. This can result in a change of the magnetic order, as for
example in \MWO\ where the $udu$ magnetic order becomes $duu$
as $A_1$ interchanges with $A_2$.  In order to arrive at the
magnetic ground state, either the spins on both $A_1$ and $A_2$
cations have to flip so that $duu$ becomes $udu$, or the spin
on the $B$ cation has to flip so that $duu$ becomes $dud$. The
former case happens at the $udu$-$udu$ DW, and the latter case
happens at the $udu$-$dud$ DW.

The above-mentioned first-principles methods can be used to predict the
DW-mediated reversal barrier in any corundum derivative, but it would
be more valuable if some intuitive rules of thumb can be summarized to
enhance our understanding. In the discussion of the polar metal \LOO, which
is also a corundum derivative, it has been pointed out that the polar
distortion is driven by short-range interactions, \cite{He_hyperLNO}
or from the crystal chemistry point of view, it is caused by the local
bonding preference of Li cations. \cite{FEmetal-Benedek} As the Li
cations are loosely bonded in the centrosymmetric structure, they
prefer a polar distortion to strengthen the local bonding environment.
Because of the structural similarity between metallic \LOO\ and other
insulating corundum derivatives, it is worth investigating
the relationship between the bonding environment of the $A$ cations and the
DW-mediated reversal barrier.

\begin{figure}
\begin{center}
  \includegraphics[width=8.5cm]{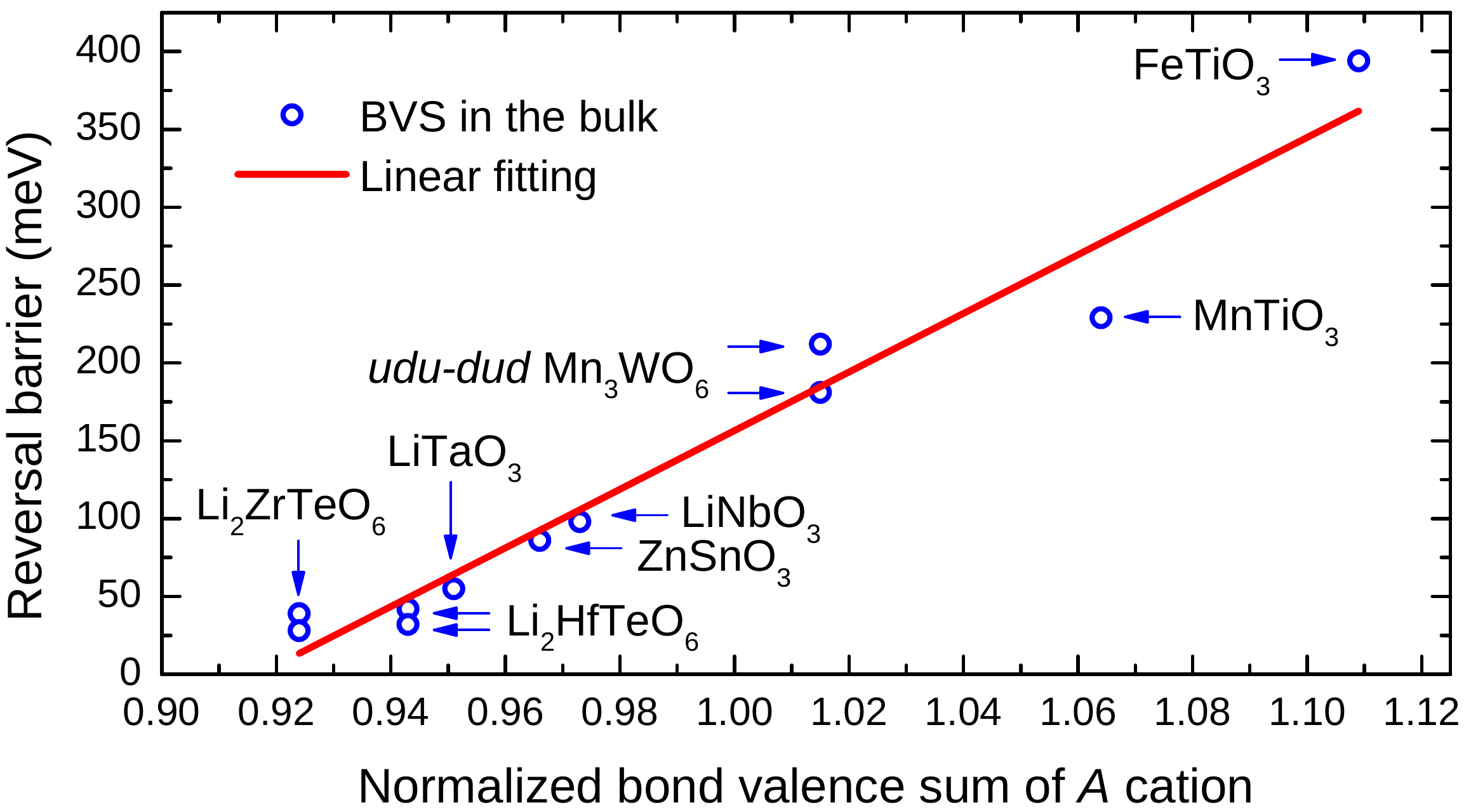}
    \caption{Scatterplot of DW-mediated reversal barrier {\it
    versus} normalized $A$-cation bond valence sum (i.e., divided
    by valence charge).  The linear fit is of the form $y=a+b(x-1)$
    with $a=157$\,meV and $b=1883$\,meV.}
    \label{fig:BVS}
\end{center}
\end{figure}
The local bonding of the $A$ cation can be described by the empirical
bond valence sum (BVS) through the equation \cite{BVS-Brown}
\beq
V_{\rm BVS}=\sum_i {\rm exp} [(R_0-R_i)/b]\,.
\label{eq:BVS}
\eeq
$V_{\rm BVS}$ estimates the number of electrons that are associated
with the local bonds. Here $R_i$ is the bond length between $A$ cations
and the $i$th nearest neighboring oxygen anions, $R_0$ is a tabulated
parameter expressing the ideal bond length when
$V_{\rm BVS}$ is exactly equal to the valence of the $A$ cation
and $b$ is an empirical constant 0.37\,\AA. For
ordered-LNO structures, the two $A$ cations are inequivalent.
Instead of using the average $V_{\rm BVS}$ of two $A$ cations, it is more
relevant to only consider the $A$ cation that is closer to the oxygen
plane, because that is the one that would migrate first in the reversal
process. Using the bond-length values extracted from bulk structures
and Eq.~(\ref{eq:BVS}), we plot the energy barrier {\it versus} the
normalized $V_{\rm BVS}$ in Fig.~\ref{fig:BVS}. A roughly linear
relationship is observed between the normalized $V_{\rm BVS}$
of the $A$ cations and the DW-mediated reversal barriers, which
also implies the dominance of short-range interactions in the
corundum derivatives. The bond valence model is also the basis of
empirical interatomic potentials that have been successfully used for
molecular dynamics simulations of ferroelectric perovskites such as
PbTiO$_3$. \cite{BVMD}

\section{Summary and outlook}
\label{sec:summary}
A DW is a topological defect that may exhibit rich physics not present
in the bulk. FE DWs separate two polar domains. In this work, we have
studied the properties of FE DWs in corundum derivatives. The mobile
180$^\circ$ charged-neutral DWs are constructed by interchanging
$A$-cation and vacancy sublattices while preserving the $B$/$B'$ and O
sublattices. Interestingly, it is found that the mobile FE domains are
interlocked with chirality domains, so that the FE DW is also a chiral
DW. For the orientation of DWs, our calculations suggest that the
Y-wall lying in the $y$-$z$ plane is more stable than the X-wall lying
in the $x$-$z$ plane.

We have also considered the magnetic order at the DW, and found that in
\MWO\ with $udu$ magnetic order, the domains with opposite polarization
also have opposite magnetization, as controlled by exchange-striction
interactions at the DW. Therefore both the polarization and magnetization
can be controlled by an electric field. We note that the magnetic
structure of \MWO\ is still under investigation,\cite{Mn3WO} and it is
possible that the actual magnetic structure has a larger magnetic cell
as in \NTO, or even a non-collinear magnetic order, instead of our
assumed collinear $udu$ magnetic order. Nevertheless, we expect that
this kind of DW-mediated magnetoelectric coupling may still exist in
the presence of a more complicated magnetic order, or in other magnetic
corundum derivates.

Since the FE polarization reversal is achieved through DW motion, we
have also studied the DW-mediated polarization reversal pathways and
barriers by applying structural constraints at the DW. We found that
the DW-mediated reversal barrier is approximately linearly correlated
with the bond valence sum of the $A$ cations. Since the local bonding
environment is accessible to experiment, this linear relationship can
provide a useful rule of thumb in predicting the DW-mediated reversal
barrier of potential new FE materials of this class.

Finally, we note that there are other classes of DWs, such as purely
chiral DWs or simultaneously chiral and magnetic DWs, that may occur
in these materials. Although not relevant for FE switching, such
domain walls might have other interesting properties that would be
worth investigating.

\section{Acknowledgment}

We thank Sang-Wook Cheong and Weida Wu for their insightful suggestions.
The work was supported by ONR grant N00014-12-1-1035.


\bibliography{mybib}

\begin{thebibliography}{28}%
\makeatletter
\providecommand \@ifxundefined [1]{%
 \@ifx{#1\undefined}
}%
\providecommand \@ifnum [1]{%
 \ifnum #1\expandafter \@firstoftwo
 \else \expandafter \@secondoftwo
 \fi
}%
\providecommand \@ifx [1]{%
 \ifx #1\expandafter \@firstoftwo
 \else \expandafter \@secondoftwo
 \fi
}%
\providecommand \natexlab [1]{#1}%
\providecommand \enquote  [1]{``#1''}%
\providecommand \bibnamefont  [1]{#1}%
\providecommand \bibfnamefont [1]{#1}%
\providecommand \citenamefont [1]{#1}%
\providecommand \href@noop [0]{\@secondoftwo}%
\providecommand \href [0]{\begingroup \@sanitize@url \@href}%
\providecommand \@href[1]{\@@startlink{#1}\@@href}%
\providecommand \@@href[1]{\endgroup#1\@@endlink}%
\providecommand \@sanitize@url [0]{\catcode `\\12\catcode `\$12\catcode
  `\&12\catcode `\#12\catcode `\^12\catcode `\_12\catcode `\%12\relax}%
\providecommand \@@startlink[1]{}%
\providecommand \@@endlink[0]{}%
\providecommand \url  [0]{\begingroup\@sanitize@url \@url }%
\providecommand \@url [1]{\endgroup\@href {#1}{\urlprefix }}%
\providecommand \urlprefix  [0]{URL }%
\providecommand \Eprint [0]{\href }%
\providecommand \doibase [0]{http://dx.doi.org/}%
\providecommand \selectlanguage [0]{\@gobble}%
\providecommand \bibinfo  [0]{\@secondoftwo}%
\providecommand \bibfield  [0]{\@secondoftwo}%
\providecommand \translation [1]{[#1]}%
\providecommand \BibitemOpen [0]{}%
\providecommand \bibitemStop [0]{}%
\providecommand \bibitemNoStop [0]{.\EOS\space}%
\providecommand \EOS [0]{\spacefactor3000\relax}%
\providecommand \BibitemShut  [1]{\csname bibitem#1\endcsname}%
\let\auto@bib@innerbib\@empty
\bibitem [{\citenamefont {Rabe}\ \emph {et~al.}(2007)\citenamefont {Rabe},
  \citenamefont {Ahn},\ and\ \citenamefont {Triscone}}]{07Book-Rabe}%
  \BibitemOpen
  \bibfield  {author} {\bibinfo {author} {\bibfnamefont {Karin~M.}\
  \bibnamefont {Rabe}}, \bibinfo {author} {\bibfnamefont {Charles~H}\
  \bibnamefont {Ahn}}, \ and\ \bibinfo {author} {\bibfnamefont {Jean-Marc}\
  \bibnamefont {Triscone}},\ }\href@noop {} {\emph {\bibinfo {title} {Physics
  of Ferroelectrics}}},\ \bibinfo {edition} {1st}\ ed.,\ Vol.\ \bibinfo
  {volume} {105}\ (\bibinfo  {publisher} {Springer-Verlag Berlin Heidelberg},\
  \bibinfo {year} {2007})\BibitemShut {NoStop}%
\bibitem [{\citenamefont {Muralt}(2000)}]{00FEsensor}%
  \BibitemOpen
  \bibfield  {author} {\bibinfo {author} {\bibfnamefont {P.}~\bibnamefont
  {Muralt}},\ }\bibfield  {title} {\enquote {\bibinfo {title} {Ferroelectric
  thin films for micro-sensors and actuators: a review},}\ }\href@noop {}
  {\bibfield  {journal} {\bibinfo  {journal} {Journal of Micromechanics and
  Microengineering}\ }\textbf {\bibinfo {volume} {10}},\ \bibinfo {pages} {136}
  (\bibinfo {year} {2000})}\BibitemShut {NoStop}%
\bibitem [{\citenamefont {Scott}(2000)}]{00Book-Scott}%
  \BibitemOpen
  \bibfield  {author} {\bibinfo {author} {\bibfnamefont {James~F.}\
  \bibnamefont {Scott}},\ }\href@noop {} {\emph {\bibinfo {title}
  {Ferroelectric Memories}}},\ \bibinfo {edition} {1st}\ ed.,\ Vol.~\bibinfo
  {volume} {3}\ (\bibinfo  {publisher} {Springer-Verlag Berlin Heidelberg},\
  \bibinfo {year} {2000})\BibitemShut {NoStop}%
\bibitem [{\citenamefont {He}\ and\ \citenamefont
  {Vanderbilt}(2003)}]{03He-defectDW}%
  \BibitemOpen
  \bibfield  {author} {\bibinfo {author} {\bibfnamefont {Lixin}\ \bibnamefont
  {He}}\ and\ \bibinfo {author} {\bibfnamefont {David}\ \bibnamefont
  {Vanderbilt}},\ }\bibfield  {title} {\enquote {\bibinfo {title}
  {First-principles study of oxygen-vacancy pinning of domain walls in
  {${\mathrm{PbTiO}}_{3}$}},}\ }\href {\doibase 10.1103/PhysRevB.68.134103}
  {\bibfield  {journal} {\bibinfo  {journal} {Phys. Rev. B}\ }\textbf {\bibinfo
  {volume} {68}},\ \bibinfo {pages} {134103} (\bibinfo {year}
  {2003})}\BibitemShut {NoStop}%
\bibitem [{\citenamefont {Seidel}\ \emph {et~al.}(2009)\citenamefont {Seidel},
  \citenamefont {Martin}, \citenamefont {He}, \citenamefont {Zhan},
  \citenamefont {Chu}, \citenamefont {Rother}, \citenamefont {Hawkridge},
  \citenamefont {Maksymovych}, \citenamefont {Yu}, \citenamefont {Gajek} \emph
  {et~al.}}]{09BFODWconduct}%
  \BibitemOpen
  \bibfield  {author} {\bibinfo {author} {\bibfnamefont {Jan}\ \bibnamefont
  {Seidel}}, \bibinfo {author} {\bibfnamefont {Lane~W}\ \bibnamefont {Martin}},
  \bibinfo {author} {\bibfnamefont {Q}~\bibnamefont {He}}, \bibinfo {author}
  {\bibfnamefont {Q}~\bibnamefont {Zhan}}, \bibinfo {author} {\bibfnamefont
  {Y-H}\ \bibnamefont {Chu}}, \bibinfo {author} {\bibfnamefont {A}~\bibnamefont
  {Rother}}, \bibinfo {author} {\bibfnamefont {ME}~\bibnamefont {Hawkridge}},
  \bibinfo {author} {\bibfnamefont {P}~\bibnamefont {Maksymovych}}, \bibinfo
  {author} {\bibfnamefont {P}~\bibnamefont {Yu}}, \bibinfo {author}
  {\bibfnamefont {M}~\bibnamefont {Gajek}},  \emph {et~al.},\ }\bibfield
  {title} {\enquote {\bibinfo {title} {Conduction at domain walls in oxide
  multiferroics},}\ }\href@noop {} {\bibfield  {journal} {\bibinfo  {journal}
  {Nature Materials}\ }\textbf {\bibinfo {volume} {8}},\ \bibinfo {pages}
  {229--234} (\bibinfo {year} {2009})}\BibitemShut {NoStop}%
\bibitem [{\citenamefont {Yang}\ \emph {et~al.}(2010)\citenamefont {Yang},
  \citenamefont {Seidel}, \citenamefont {Byrnes}, \citenamefont {Shafer},
  \citenamefont {Yang}, \citenamefont {Rossell}, \citenamefont {Yu},
  \citenamefont {Chu}, \citenamefont {Scott}, \citenamefont {Ager} \emph
  {et~al.}}]{10BFODWPV}%
  \BibitemOpen
  \bibfield  {author} {\bibinfo {author} {\bibfnamefont {SY}~\bibnamefont
  {Yang}}, \bibinfo {author} {\bibfnamefont {J}~\bibnamefont {Seidel}},
  \bibinfo {author} {\bibfnamefont {SJ}~\bibnamefont {Byrnes}}, \bibinfo
  {author} {\bibfnamefont {P}~\bibnamefont {Shafer}}, \bibinfo {author}
  {\bibfnamefont {C-H}\ \bibnamefont {Yang}}, \bibinfo {author} {\bibfnamefont
  {MD}~\bibnamefont {Rossell}}, \bibinfo {author} {\bibfnamefont
  {P}~\bibnamefont {Yu}}, \bibinfo {author} {\bibfnamefont {Y-H}\ \bibnamefont
  {Chu}}, \bibinfo {author} {\bibfnamefont {JF}~\bibnamefont {Scott}}, \bibinfo
  {author} {\bibfnamefont {JW}~\bibnamefont {Ager}},  \emph {et~al.},\
  }\bibfield  {title} {\enquote {\bibinfo {title} {Above-bandgap voltages from
  ferroelectric photovoltaic devices},}\ }\href@noop {} {\bibfield  {journal}
  {\bibinfo  {journal} {Nature nanotechnology}\ }\textbf {\bibinfo {volume}
  {5}},\ \bibinfo {pages} {143--147} (\bibinfo {year} {2010})}\BibitemShut
  {NoStop}%
\bibitem [{\citenamefont {Oh}\ \emph {et~al.}(2015)\citenamefont {Oh},
  \citenamefont {Luo}, \citenamefont {Huang}, \citenamefont {Wang},\ and\
  \citenamefont {Cheong}}]{15chargedDW}%
  \BibitemOpen
  \bibfield  {author} {\bibinfo {author} {\bibfnamefont {Yoon~Seok}\
  \bibnamefont {Oh}}, \bibinfo {author} {\bibfnamefont {Xuan}\ \bibnamefont
  {Luo}}, \bibinfo {author} {\bibfnamefont {Fei-Ting}\ \bibnamefont {Huang}},
  \bibinfo {author} {\bibfnamefont {Yazhong}\ \bibnamefont {Wang}}, \ and\
  \bibinfo {author} {\bibfnamefont {Sang-Wook}\ \bibnamefont {Cheong}},\
  }\bibfield  {title} {\enquote {\bibinfo {title} {Experimental demonstration
  of hybrid improper ferroelectricity and the presence of abundant charged
  walls in {(Ca,Sr)$_3$Ti$_2$O$_7$} crystals},}\ }\href@noop {} {\bibfield
  {journal} {\bibinfo  {journal} {Nature Materials}\ }\textbf {\bibinfo
  {volume} {14}},\ \bibinfo {pages} {407--413} (\bibinfo {year}
  {2015})}\BibitemShut {NoStop}%
\bibitem [{\citenamefont {Choi}\ \emph {et~al.}(2010)\citenamefont {Choi},
  \citenamefont {Horibe}, \citenamefont {Yi}, \citenamefont {Choi},
  \citenamefont {Wu},\ and\ \citenamefont {Cheong}}]{10YMnO3FEvortex}%
  \BibitemOpen
  \bibfield  {author} {\bibinfo {author} {\bibfnamefont {T}~\bibnamefont
  {Choi}}, \bibinfo {author} {\bibfnamefont {Y}~\bibnamefont {Horibe}},
  \bibinfo {author} {\bibfnamefont {HT}~\bibnamefont {Yi}}, \bibinfo {author}
  {\bibfnamefont {YJ}~\bibnamefont {Choi}}, \bibinfo {author} {\bibfnamefont
  {Weida}\ \bibnamefont {Wu}}, \ and\ \bibinfo {author} {\bibfnamefont {S-W}\
  \bibnamefont {Cheong}},\ }\bibfield  {title} {\enquote {\bibinfo {title}
  {Insulating interlocked ferroelectric and structural antiphase domain walls
  in multiferroic {YMnO$_3$}},}\ }\href@noop {} {\bibfield  {journal} {\bibinfo
   {journal} {Nature Materials}\ }\textbf {\bibinfo {volume} {9}},\ \bibinfo
  {pages} {253--258} (\bibinfo {year} {2010})}\BibitemShut {NoStop}%
\bibitem [{\citenamefont {Geng}\ \emph {et~al.}(2012)\citenamefont {Geng},
  \citenamefont {Lee}, \citenamefont {Choi}, \citenamefont {Cheong},\ and\
  \citenamefont {Wu}}]{12RMnOMFvortex}%
  \BibitemOpen
  \bibfield  {author} {\bibinfo {author} {\bibfnamefont {Yanan}\ \bibnamefont
  {Geng}}, \bibinfo {author} {\bibfnamefont {Nara}\ \bibnamefont {Lee}},
  \bibinfo {author} {\bibfnamefont {YJ}~\bibnamefont {Choi}}, \bibinfo {author}
  {\bibfnamefont {S-W}\ \bibnamefont {Cheong}}, \ and\ \bibinfo {author}
  {\bibfnamefont {Weida}\ \bibnamefont {Wu}},\ }\bibfield  {title} {\enquote
  {\bibinfo {title} {Collective magnetism at multiferroic vortex domain
  walls},}\ }\href@noop {} {\bibfield  {journal} {\bibinfo  {journal} {Nano
  Letters}\ }\textbf {\bibinfo {volume} {12}},\ \bibinfo {pages} {6055--6059}
  (\bibinfo {year} {2012})}\BibitemShut {NoStop}%
\bibitem [{\citenamefont {Wang}\ \emph {et~al.}(2015)\citenamefont {Wang},
  \citenamefont {Huang}, \citenamefont {Yang}, \citenamefont {Oh},\ and\
  \citenamefont {Cheong}}]{15NTOchiralFEDW}%
  \BibitemOpen
  \bibfield  {author} {\bibinfo {author} {\bibfnamefont {Xueyun}\ \bibnamefont
  {Wang}}, \bibinfo {author} {\bibfnamefont {Fei-Ting}\ \bibnamefont {Huang}},
  \bibinfo {author} {\bibfnamefont {Junjie}\ \bibnamefont {Yang}}, \bibinfo
  {author} {\bibfnamefont {Yoon~Seok}\ \bibnamefont {Oh}}, \ and\ \bibinfo
  {author} {\bibfnamefont {Sang-Wook}\ \bibnamefont {Cheong}},\ }\bibfield
  {title} {\enquote {\bibinfo {title} {Interlocked chiral/polar domain walls
  and large optical rotation in {Ni$_3$TeO$_6$}},}\ }\href@noop {} {\bibfield
  {journal} {\bibinfo  {journal} {APL Materials}\ }\textbf {\bibinfo {volume}
  {3}},\ \bibinfo {pages} {076105} (\bibinfo {year} {2015})}\BibitemShut
  {NoStop}%
\bibitem [{\citenamefont {Ye}\ and\ \citenamefont
  {Vanderbilt}(2016)}]{MY16-FEcd}%
  \BibitemOpen
  \bibfield  {author} {\bibinfo {author} {\bibfnamefont {Meng}\ \bibnamefont
  {Ye}}\ and\ \bibinfo {author} {\bibfnamefont {David}\ \bibnamefont
  {Vanderbilt}},\ }\bibfield  {title} {\enquote {\bibinfo {title}
  {Ferroelectricity in corundum derivatives},}\ }\href {\doibase
  10.1103/PhysRevB.93.134303} {\bibfield  {journal} {\bibinfo  {journal} {Phys.
  Rev. B}\ }\textbf {\bibinfo {volume} {93}},\ \bibinfo {pages} {134303}
  (\bibinfo {year} {2016})}\BibitemShut {NoStop}%
\bibitem [{\citenamefont {Kresse}\ and\ \citenamefont
  {Furthm\"uller}(1996)}]{VASP}%
  \BibitemOpen
  \bibfield  {author} {\bibinfo {author} {\bibfnamefont {G.}~\bibnamefont
  {Kresse}}\ and\ \bibinfo {author} {\bibfnamefont {J.}~\bibnamefont
  {Furthm\"uller}},\ }\bibfield  {title} {\enquote {\bibinfo {title} {Efficient
  iterative schemes for \textit{ab initio} total-energy calculations using a
  plane-wave basis set},}\ }\href {\doibase 10.1103/PhysRevB.54.11169}
  {\bibfield  {journal} {\bibinfo  {journal} {Phys. Rev. B}\ }\textbf {\bibinfo
  {volume} {54}},\ \bibinfo {pages} {11169--11186} (\bibinfo {year}
  {1996})}\BibitemShut {NoStop}%
\bibitem [{\citenamefont {Perdew}\ \emph {et~al.}(2008)\citenamefont {Perdew},
  \citenamefont {Ruzsinszky}, \citenamefont {Csonka}, \citenamefont {Vydrov},
  \citenamefont {Scuseria}, \citenamefont {Constantin}, \citenamefont {Zhou},\
  and\ \citenamefont {Burke}}]{PBEsol}%
  \BibitemOpen
  \bibfield  {author} {\bibinfo {author} {\bibfnamefont {John~P.}\ \bibnamefont
  {Perdew}}, \bibinfo {author} {\bibfnamefont {Adrienn}\ \bibnamefont
  {Ruzsinszky}}, \bibinfo {author} {\bibfnamefont {G\'abor~I.}\ \bibnamefont
  {Csonka}}, \bibinfo {author} {\bibfnamefont {Oleg~A.}\ \bibnamefont
  {Vydrov}}, \bibinfo {author} {\bibfnamefont {Gustavo~E.}\ \bibnamefont
  {Scuseria}}, \bibinfo {author} {\bibfnamefont {Lucian~A.}\ \bibnamefont
  {Constantin}}, \bibinfo {author} {\bibfnamefont {Xiaolan}\ \bibnamefont
  {Zhou}}, \ and\ \bibinfo {author} {\bibfnamefont {Kieron}\ \bibnamefont
  {Burke}},\ }\bibfield  {title} {\enquote {\bibinfo {title} {Restoring the
  density-gradient expansion for exchange in solids and surfaces},}\ }\href
  {\doibase 10.1103/PhysRevLett.100.136406} {\bibfield  {journal} {\bibinfo
  {journal} {Phys. Rev. Lett.}\ }\textbf {\bibinfo {volume} {100}},\ \bibinfo
  {pages} {136406} (\bibinfo {year} {2008})}\BibitemShut {NoStop}%
\bibitem [{\citenamefont {Bl\"ochl}(1994)}]{PAW}%
  \BibitemOpen
  \bibfield  {author} {\bibinfo {author} {\bibfnamefont {P.~E.}\ \bibnamefont
  {Bl\"ochl}},\ }\bibfield  {title} {\enquote {\bibinfo {title} {Projector
  augmented-wave method},}\ }\href {\doibase 10.1103/PhysRevB.50.17953}
  {\bibfield  {journal} {\bibinfo  {journal} {Phys. Rev. B}\ }\textbf {\bibinfo
  {volume} {50}},\ \bibinfo {pages} {17953--17979} (\bibinfo {year}
  {1994})}\BibitemShut {NoStop}%
\bibitem [{\citenamefont {Dudarev}\ \emph {et~al.}(1998)\citenamefont
  {Dudarev}, \citenamefont {Botton}, \citenamefont {Savrasov}, \citenamefont
  {Humphreys},\ and\ \citenamefont {Sutton}}]{LDAU}%
  \BibitemOpen
  \bibfield  {author} {\bibinfo {author} {\bibfnamefont {S.~L.}\ \bibnamefont
  {Dudarev}}, \bibinfo {author} {\bibfnamefont {G.~A.}\ \bibnamefont {Botton}},
  \bibinfo {author} {\bibfnamefont {S.~Y.}\ \bibnamefont {Savrasov}}, \bibinfo
  {author} {\bibfnamefont {C.~J.}\ \bibnamefont {Humphreys}}, \ and\ \bibinfo
  {author} {\bibfnamefont {A.~P.}\ \bibnamefont {Sutton}},\ }\bibfield  {title}
  {\enquote {\bibinfo {title} {Electron-energy-loss spectra and the structural
  stability of nickel oxide: {An} {LSDA+U} study},}\ }\href {\doibase
  10.1103/PhysRevB.57.1505} {\bibfield  {journal} {\bibinfo  {journal} {Phys.
  Rev. B}\ }\textbf {\bibinfo {volume} {57}},\ \bibinfo {pages} {1505--1509}
  (\bibinfo {year} {1998})}\BibitemShut {NoStop}%
\bibitem [{\citenamefont {Inbar}\ and\ \citenamefont
  {Cohen}(1996)}]{LiNb/TaO-cal}%
  \BibitemOpen
  \bibfield  {author} {\bibinfo {author} {\bibfnamefont {Iris}\ \bibnamefont
  {Inbar}}\ and\ \bibinfo {author} {\bibfnamefont {R.~E.}\ \bibnamefont
  {Cohen}},\ }\bibfield  {title} {\enquote {\bibinfo {title} {Comparison of the
  electronic structures and energetics of ferroelectric
  {${\mathrm{LiNbO}}_{3}$} and {${\mathrm{LiTaO}}_{3}$}},}\ }\href {\doibase
  10.1103/PhysRevB.53.1193} {\bibfield  {journal} {\bibinfo  {journal} {Phys.
  Rev. B}\ }\textbf {\bibinfo {volume} {53}},\ \bibinfo {pages} {1193--1204}
  (\bibinfo {year} {1996})}\BibitemShut {NoStop}%
\bibitem [{\citenamefont {Veithen}\ and\ \citenamefont
  {Ghosez}(2002)}]{LiNb/TaO-cal2}%
  \BibitemOpen
  \bibfield  {author} {\bibinfo {author} {\bibfnamefont {M.}~\bibnamefont
  {Veithen}}\ and\ \bibinfo {author} {\bibfnamefont {Ph.}\ \bibnamefont
  {Ghosez}},\ }\bibfield  {title} {\enquote {\bibinfo {title} {First-principles
  study of the dielectric and dynamical properties of lithium niobate},}\
  }\href@noop {} {\bibfield  {journal} {\bibinfo  {journal} {Phys. Rev. B}\
  }\textbf {\bibinfo {volume} {65}},\ \bibinfo {pages} {214302} (\bibinfo
  {year} {2002})}\BibitemShut {NoStop}%
\bibitem [{\citenamefont {Groth}(1893)}]{Wulff}%
  \BibitemOpen
  \bibfield  {author} {\bibinfo {author} {\bibfnamefont {Paul}\ \bibnamefont
  {Groth}},\ }\href@noop {} {\emph {\bibinfo {title} {Zeitschrift f{\"u}r
  Krystallographie und Mineralogie}}},\ Vol.~\bibinfo {volume} {11}\ (\bibinfo
  {publisher} {Wilhelm Engelmann},\ \bibinfo {year} {1893})\BibitemShut
  {NoStop}%
\bibitem [{\citenamefont {Lee}\ \emph {et~al.}(2011)\citenamefont {Lee},
  \citenamefont {Xu}, \citenamefont {Dierolf}, \citenamefont {Gopalan},\ and\
  \citenamefont {Phillpot}}]{LiNbO-DWexpt}%
  \BibitemOpen
  \bibfield  {author} {\bibinfo {author} {\bibfnamefont {Donghwa}\ \bibnamefont
  {Lee}}, \bibinfo {author} {\bibfnamefont {Haixuan}\ \bibnamefont {Xu}},
  \bibinfo {author} {\bibfnamefont {Volkmar}\ \bibnamefont {Dierolf}}, \bibinfo
  {author} {\bibfnamefont {Venkatraman}\ \bibnamefont {Gopalan}}, \ and\
  \bibinfo {author} {\bibfnamefont {Simon~R.}\ \bibnamefont {Phillpot}},\
  }\bibfield  {title} {\enquote {\bibinfo {title} {Shape of ferroelectric
  domains in {LiNbO$_3$} and {LiTaO$_3$} from defect/domain-wall
  interactions},}\ }\href {\doibase 10.1063/1.3560343} {\bibfield  {journal}
  {\bibinfo  {journal} {Applied Physics Letters}\ }\textbf {\bibinfo {volume}
  {98}},\ \bibinfo {eid} {092903} (\bibinfo {year} {2011}),\
  10.1063/1.3560343}\BibitemShut {NoStop}%
\bibitem [{\citenamefont {Padilla}\ \emph {et~al.}(1996)\citenamefont
  {Padilla}, \citenamefont {Zhong},\ and\ \citenamefont
  {Vanderbilt}}]{96BTODW}%
  \BibitemOpen
  \bibfield  {author} {\bibinfo {author} {\bibfnamefont {J.}~\bibnamefont
  {Padilla}}, \bibinfo {author} {\bibfnamefont {W.}~\bibnamefont {Zhong}}, \
  and\ \bibinfo {author} {\bibfnamefont {David}\ \bibnamefont {Vanderbilt}},\
  }\bibfield  {title} {\enquote {\bibinfo {title} {First-principles
  investigation of $180^{\circ}$ domain walls in {BaTi${\mathrm{O}}_{3}$}},}\
  }\href {\doibase 10.1103/PhysRevB.53.R5969} {\bibfield  {journal} {\bibinfo
  {journal} {Phys. Rev. B}\ }\textbf {\bibinfo {volume} {53}},\ \bibinfo
  {pages} {R5969--R5973} (\bibinfo {year} {1996})}\BibitemShut {NoStop}%
\bibitem [{\citenamefont {Meyer}\ and\ \citenamefont
  {Vanderbilt}(2002)}]{02PTODW}%
  \BibitemOpen
  \bibfield  {author} {\bibinfo {author} {\bibfnamefont {B.}~\bibnamefont
  {Meyer}}\ and\ \bibinfo {author} {\bibfnamefont {David}\ \bibnamefont
  {Vanderbilt}},\ }\bibfield  {title} {\enquote {\bibinfo {title} {\textit{Ab
  initio} study of ferroelectric domain walls in {${\mathrm{PbTiO}}_{3}$}},}\
  }\href {\doibase 10.1103/PhysRevB.65.104111} {\bibfield  {journal} {\bibinfo
  {journal} {Phys. Rev. B}\ }\textbf {\bibinfo {volume} {65}},\ \bibinfo
  {pages} {104111} (\bibinfo {year} {2002})}\BibitemShut {NoStop}%
\bibitem [{\citenamefont {Lee}\ \emph {et~al.}(2010)\citenamefont {Lee},
  \citenamefont {Xu}, \citenamefont {Dierolf}, \citenamefont {Gopalan},\ and\
  \citenamefont {Phillpot}}]{10LNODWcal}%
  \BibitemOpen
  \bibfield  {author} {\bibinfo {author} {\bibfnamefont {Donghwa}\ \bibnamefont
  {Lee}}, \bibinfo {author} {\bibfnamefont {Haixuan}\ \bibnamefont {Xu}},
  \bibinfo {author} {\bibfnamefont {Volkmar}\ \bibnamefont {Dierolf}}, \bibinfo
  {author} {\bibfnamefont {Venkatraman}\ \bibnamefont {Gopalan}}, \ and\
  \bibinfo {author} {\bibfnamefont {Simon~R.}\ \bibnamefont {Phillpot}},\
  }\bibfield  {title} {\enquote {\bibinfo {title} {Structure and energetics of
  ferroelectric domain walls in {${\text{LiNbO}}_{3}$} from atomic-level
  simulations},}\ }\href {\doibase 10.1103/PhysRevB.82.014104} {\bibfield
  {journal} {\bibinfo  {journal} {Phys. Rev. B}\ }\textbf {\bibinfo {volume}
  {82}},\ \bibinfo {pages} {014104} (\bibinfo {year} {2010})}\BibitemShut
  {NoStop}%
\bibitem [{\citenamefont {Benedek}\ and\ \citenamefont
  {Fennie}(2011)}]{11Benedek-HybridImproper}%
  \BibitemOpen
  \bibfield  {author} {\bibinfo {author} {\bibfnamefont {Nicole~A.}\
  \bibnamefont {Benedek}}\ and\ \bibinfo {author} {\bibfnamefont {Craig~J.}\
  \bibnamefont {Fennie}},\ }\bibfield  {title} {\enquote {\bibinfo {title}
  {Hybrid improper ferroelectricity: {A} mechanism for controllable
  polarization-magnetization coupling},}\ }\href {\doibase
  10.1103/PhysRevLett.106.107204} {\bibfield  {journal} {\bibinfo  {journal}
  {Phys. Rev. Lett.}\ }\textbf {\bibinfo {volume} {106}},\ \bibinfo {pages}
  {107204} (\bibinfo {year} {2011})}\BibitemShut {NoStop}%
\bibitem [{\citenamefont {Li}\ \emph {et~al.}(2016)\citenamefont {Li},
  \citenamefont {Ren}, \citenamefont {Guo},\ and\ \citenamefont
  {He}}]{He_hyperLNO}%
  \BibitemOpen
  \bibfield  {author} {\bibinfo {author} {\bibfnamefont {Pengfei}\ \bibnamefont
  {Li}}, \bibinfo {author} {\bibfnamefont {Xinguo}\ \bibnamefont {Ren}},
  \bibinfo {author} {\bibfnamefont {Guang-Can}\ \bibnamefont {Guo}}, \ and\
  \bibinfo {author} {\bibfnamefont {Lixin}\ \bibnamefont {He}},\ }\bibfield
  {title} {\enquote {\bibinfo {title} {The origin of hyperferroelectricity in
  {Li$B$O$_3$} {($B$ = V, Nb, Ta, Os)}},}\ }\href {\doibase 10.1038/srep34085}
  {\bibfield  {journal} {\bibinfo  {journal} {Scientific Reports}\ }\textbf
  {\bibinfo {volume} {6}},\ \bibinfo {pages} {34085} (\bibinfo {year}
  {2016})}\BibitemShut {NoStop}%
\bibitem [{\citenamefont {Benedek}\ and\ \citenamefont
  {Birol}(2016)}]{FEmetal-Benedek}%
  \BibitemOpen
  \bibfield  {author} {\bibinfo {author} {\bibfnamefont {Nicole~A}\
  \bibnamefont {Benedek}}\ and\ \bibinfo {author} {\bibfnamefont {Turan}\
  \bibnamefont {Birol}},\ }\bibfield  {title} {\enquote {\bibinfo {title}
  {{`Ferroelectric'} metals reexamined: {Fundamental} mechanisms and design
  considerations for new materials},}\ }\href@noop {} {\bibfield  {journal}
  {\bibinfo  {journal} {Journal of Materials Chemistry C}\ }\textbf {\bibinfo
  {volume} {4}},\ \bibinfo {pages} {4000--4015} (\bibinfo {year}
  {2016})}\BibitemShut {NoStop}%
\bibitem [{\citenamefont {Brown}\ and\ \citenamefont
  {Altermatt}(1985)}]{BVS-Brown}%
  \BibitemOpen
  \bibfield  {author} {\bibinfo {author} {\bibfnamefont {I.~D.}\ \bibnamefont
  {Brown}}\ and\ \bibinfo {author} {\bibfnamefont {D.}~\bibnamefont
  {Altermatt}},\ }\bibfield  {title} {\enquote {\bibinfo {title} {{Bond-valence
  parameters obtained from a systematic analysis of the {Inorganic} {Crystal}
  {Structure} {Database}}},}\ }\href {\doibase 10.1107/S0108768185002063}
  {\bibfield  {journal} {\bibinfo  {journal} {Acta Crystallographica Section
  B}\ }\textbf {\bibinfo {volume} {41}},\ \bibinfo {pages} {244--247} (\bibinfo
  {year} {1985})}\BibitemShut {NoStop}%
\bibitem [{\citenamefont {Shin}\ \emph {et~al.}(2005)\citenamefont {Shin},
  \citenamefont {Cooper}, \citenamefont {Grinberg},\ and\ \citenamefont
  {Rappe}}]{BVMD}%
  \BibitemOpen
  \bibfield  {author} {\bibinfo {author} {\bibfnamefont {Young-Han}\
  \bibnamefont {Shin}}, \bibinfo {author} {\bibfnamefont {Valentino~R.}\
  \bibnamefont {Cooper}}, \bibinfo {author} {\bibfnamefont {Ilya}\ \bibnamefont
  {Grinberg}}, \ and\ \bibinfo {author} {\bibfnamefont {Andrew~M.}\
  \bibnamefont {Rappe}},\ }\bibfield  {title} {\enquote {\bibinfo {title}
  {Development of a bond-valence molecular-dynamics model for complex
  oxides},}\ }\href {\doibase 10.1103/PhysRevB.71.054104} {\bibfield  {journal}
  {\bibinfo  {journal} {Phys. Rev. B}\ }\textbf {\bibinfo {volume} {71}},\
  \bibinfo {pages} {054104} (\bibinfo {year} {2005})}\BibitemShut {NoStop}%
\bibitem [{Mn3()}]{Mn3WO}%
  \BibitemOpen
  \href@noop {} {}\bibinfo {note} {Private communication with Prof. Martha
  Greenblatt and Dr. Manrong Li}\BibitemShut {NoStop}%
\end{thebibliography}%
\end{document}